\documentclass[aps,showpacs,amsmath,amssymb,onecolumn,superscriptaddress,notitlepage,longbibliography,preprintnumbers,prx]{revtex4-2}

\usepackage{float}
\usepackage{comment}
\usepackage[T1]{fontenc}
\usepackage[utf8]{inputenc}
\usepackage{lmodern}
\usepackage{amsmath,amssymb,amsthm,mathtools,bm,mathrsfs}
\usepackage{bbm}
\usepackage{microtype}
\usepackage{xcolor}
\usepackage{enumitem}
\usepackage{graphicx}
\usepackage[colorlinks=true, allcolors=blue]{hyperref}
\usepackage[table, dvipsnames]{xcolor}
\usepackage{comment}

\usepackage{subcaption}
\usepackage{cleveref}
\usepackage{dsfont}
\usepackage{multirow}
\usepackage{booktabs}
\usepackage{orcidlink}
\usepackage[normalem]{ulem}
\usepackage{ragged2e}
\usepackage{calc}
\usepackage{algpseudocode}
\algrenewcommand\algorithmicrequire{\textbf{Input:}}
\algrenewcommand\algorithmicensure{\textbf{Output:}}

\makeatletter
\newcounter{algorithm}

\makeatother

\newcommand{\ket}[1]{\left|#1\right\rangle}

\begin{document}

\title{Designing Quantum Solutions to The Closest String Problem with the Non-Variational QWOA}

\author{Benjamin Rossdeutscher}
\affiliation{Centre for Quantum Information,
Simulation and Algorithms, \\ 
The University of Western Australia, Perth, Australia}
\author{Jingbo Wang}
\email{jingbo.wang@uwa.edu.au}
\affiliation{Centre for Quantum Information,
Simulation and Algorithms, \\ 
The University of Western Australia, Perth, Australia}

\date{today}

\begin{abstract}
Quantum computing offers promising approaches to combinatorial optimization problems whose search spaces grow exponentially with problem size. The Closest String problem belongs to this category, which seeks a string that minimizes the maximum Hamming distance to any string in a given collection of reference strings. In this work, we present an end-to-end quantum implementation of the Closest String problem using the Non-Variational Quantum Walk Optimization Algorithm (NV-QWOA), evaluating its resource requirements and solution amplification capabilities. Our automated workflow converts arbitrary problem instances into quantum circuits expressed in one- or two-qubit gates that encode candidate solutions, evaluate the objective function, and apply NV-QWOA optimisation to increase the measurement probabilities of high-quality solutions. Analytical resource estimates yield a qubit complexity of $\mathcal{O}(nm\log k)$ and a circuit depth of $\mathcal{O}(pm\log^2 n + pn\log n)$, where $n$ is the string length, $m$ is the number of reference strings, $k$ is the alphabet size, and $p$ is the number of NV-QWOA iterations. We introduce a revised prescription for the phase-separation parameter, $\gamma$, that depends on problem size but requires no instance-specific optimization. This prescription mitigates excessive phase wrapping modulo $2\pi$, which can weaken the relationship between objective values and the phases guiding the optimization dynamics. Across all simulated problem sizes, the algorithm consistently amplifies optimal and near-optimal candidates. This work provides an explicit gate-level implementation, an improved phase-scaling method, and a foundation for hardware experiments and benchmarking against classical algorithms.
\end{abstract}
\maketitle

\section{Introduction}

Beyond verifying the applicability of the NV-QWOA to a new problem type, the Closest String problem is, in itself worthy of a quantum solution, due to its wide applicability and classical difficulty to solve exactly. 
In terms of computational complexity, the problem is classified as NP-Hard ~\cite{CSP}, meaning that it is not classically simple to compute (there are no known exact solutions with polynomial or better time scaling), creating a demand for better and quantum solutions. 
This problem has useful applications in a diverse set of fields, from communications (in reconstructing a message or signal from multiple partially damaged copies of that signal), in chemistry (for detecting common structures in protein molecules), in genetics (for example in rapidly responding to a disease outbreak by comparing mutated strains to find the common bacteria or virus to combat or in phylogenetic research), and in code correction (when looking for copy errors across repeated code or databases), to name a few possibilities.

Hamming distance ($d_{H}$) is a count of the number of characters that need to be changed to transform one string into another, such that changing any one character in a string to any other character within the defined alphabet would create a Hamming Distance distance of 1 between the original and modified string. 
There exist both decision and combinatorial optimisation versions of these problems, where the decision problem (NP class) aims to detect whether a solution that has a cost value below a chosen threshold exists for a given problem instance, whereas the combinatorial optimisation problem seeks to find the optimal solution(s). 
We directed our research towards solving the combinatorial optimisation problem, as it sees more direct applications and is more effectively represented in its generalised, non-problem-instance specific form in a compatible manner for the NV-QWOA, as the NV-QWOA is a combinatorial optimisation algorithm. 
The formulation of the problem used for our implementation, is defined as such:
Given a set of $m$ reference strings { $s_{1}, s_{2}, ..., s_{m}
$ }, each of the length $n$ (number of characters), we are attempting to find the solution string $s$, of equal length $n$, which when input as the argument to the Closest String cost function minimise the objective function output. 
The cost function of a solution string is defined as the maximum of the values among the hamming distances taken between that solution string and each one of the input strings in the reference set~~\cite{CSP}:
\begin{equation} 
\text{f}(s)=max_{i=1,2,...,m}d_{H}(s,s_{i}).
 \end{equation} 
When classically calculating the quality matrix $Q$, for $m$ reference strings of length $n$ it is required to do $mn$ checks calculating if two characters have an equal value to calculate the cost of one potential solution string. For one solution string, where each character is drawing from an alphabet of size $k$, there are $k^n$ different possible solution strings in the solution space, meaning that classically calculating $Q$ requires $k^nmn$ character comparisons. Given $k>1$, this at minimum scales to $2^nmn$ comparisons to find $Q$ classically. As this scales exponentially with string length, it becomes unmanageable to classically calculate with even the best supercomputers for sufficiently large $n$. Finding the global minimum of the cost function through an exhaustive classical search scales as $\mathcal{O}(k^n)$ evaluations for a candidate string of length $n$ over an alphabet of size $k$. In contrast, our quantum circuit requires qubit resources on the order of $\mathcal{O}(nm \log(k))$ to encode instances of the Closest String and Substring problems. By evaluating the objective functions in superposition across the candidate space, the quantum walk dynamics concentrate measurement probabilities on optimal or near-optimal solutions, offering a favorable scaling trajectory compared to exhaustive classical enumeration. 

There also exists an alternative formulation, which has been referred to as the Consensus String Problem based on Consensus Error~~\cite{constring}, which defines the cost function of a solution string as the sum of Hamming distances between that solution string and each of the reference strings, rather than the maximum of those Hamming Distances, with the decision version of this problem being NP-complete and the optimisation version being NP-Hard. This formulation was used by Dissanayake~\cite{QUBO} when re-framing the problem into a QUBO problem to implement through quantum annealing hardware.
The objective function of the more general Closest Substring Problem is then defined as, for solution substrings of length $n$, and reference strings of length $L$, where $n\leq L$: taking the minimum of the Hamming distances between the solution substring $s$ and each substring of length $n$ within one reference string, then taking the maximum of this list across every reference string as the cost of the solution substring,
making the Closest String Problem a special case of the Closest Substring Problem, which sets the reference string length $L$ and the solution substring length $n$ to be equal.
Our research focused on solving the Closest String Problem, with generalisation to the Closest Substring Problem being a potential future extension of the research.

While not directly attempting to solve the Closest String Problem, Prousalis et al.~~\cite{stringmatch} provided an explicit quantum solution to finding areas of overlap between two strings of uneven length. One avenue of further inquiry might be into applying some of their work into the generalisation of the Closest String cost function into implementing the Closest Substring cost function.
Finally, while not a specifically intended goal of the project, there do exist other variations of the Closest String Problem, such as the "Distinguishing Substring Problem", and further generalising the our method to solve these could also be considered for future avenues of research.

\section{Literature Review}
Computational complexity classes, such as "NP" or "NP-Hard", are used to describe the difficulty of a computational problem. They are primarily based on how the length of time required to solve a problem, and the amount of time to verify a solution to a problem, scale with the problem parameters~~\cite{complexity}.
NP problems are "Nondeterministic-Polynomial time" problems. A problem is in NP if it is a decision problem (ie. it requires finding whether a valid solution exists, not what it is) and requires that any valid solution can be verified as a true valid solution in polynomial time.
NP-Complete  problems are decision problems that are in NP and are also as hard as any NP problem, in the sense that any NP problem can be reduced or represented in the form of this problem. 
NP-Hard problems are problems which are at necessarily at least as difficult as NP-complete problems, but don't necessarily meet the criteria to be an NP-complete problem (ie. a problem might be an optimisation problem, not a decision problem, so it can't be NP even if runtime scales exponentially with problem size). For a problem to be NP-Hard, all problems in NP need to be able to be reduced to it in polynomial time.
BPP (Bounded Error Probabilistic Polynomial time) problems are decision problems that can be solved within a set error margin by a classical computer (generally with a success rate of two thirds), running an algorithm containing randomised elements, in polynomial time.
BQP (Bounded Error Quantum Polynomial time) problems (similar the classical BPP), refer to decision problems, but differ in that a problem is BQP if a quantum computer can solve it within a fixed error rate of $\leq\frac{1}{3}$.
In all of these cases, solving a problem refers to finding an exact solution, in the case of optimisation problems this would require an algorithm to exactly finding the optimal solution every time (as such approximate algorithms aren't considered to have "solved" a NP-Hard problem in a way meaningful to relating complexity classes even if they produce an optimal solution the an absolute majority of the time), whereas BPP and BQP, while accepting a fixed error rate, are for decision problems.~\cite{fakhravar2022combiningheuristicsexactalgorithms}

Quantum circuits are a method of representing a  system's evolution through a gate-based representation of the transformations being applied to certain qubits in the system, ordered by the timeline of the wavefunction's evolution and when they act on it. Some of these gates have corresponding classical logic gates, such as the Pauli-X gate and a classical NOT gate, while others have no classical equivalent, such as the Hadamard gate commonly used to introduce superposition. There exist several universal sets of gates for a quantum computer, akin to classical universal sets of gates~\cite{QML}. 
A register refers to a grouping of qubits in a quantum computer dedicated for a certain role (ie. a solution register might store the superposition of the solutions, an ancilla register could be used for temporary storage of information to avoid overwriting something useful, etc.).
As all gate operations need to be reversible on a quantum computer, implementing actions such as comparisons require ancilla qubits for working space to do calculations and store the results of comparison, for cases such as needing to store results and being unable to do so by overwriting information used to make a comparison or calculation with the output of that task, as that would normally be a non-reversible operation, making it non-Unitary, and a non-viable gate, creating the need for the assistance of said ancilla qubits and registers.
Uncomputing is the practice of applying the inverse of a gate or the inverse of a series of gates (any series of unitaries can be written as one composite unitary by writing empty steps of the circuit as identity gates and finding the tensor product of all simultaneous unitaries). By applying the inverse of these unitaries to uncompute the ancilla register, it disentangles the ancilla register from the solution states and prevents them from interfering with later steps, along with freeing the space for later repeated usage.

\subsection{Quantum Algorithms}
There are other paradigms of quantum computing, such as adiabatic computing, but we used the gate-based model for all the work done here.
Quantum algorithms describe intentional ordered use of unitary transformations to manipulate a quantum state to encode information, evolve it over time then extract information from the evolved state. Hybrid quantum-classical methods involve evolving a state, making certain measurements, then using the obtained information to continue evolving that state or a new state. 
Some of the more famous quantum algorithms include: Grover's Search Algorithm which obtains a quadratic time speedup from its classical equivalents and is the optimal quantum search on a completely unstructured problem instance~\cite{grover1996fastquantummechanicalalgorithm}, Shor's factorisation algorithm which can find factors for classically intractably large numbers~\cite{Shor_1997} , Quantum Phase Estimation which can extract phase information from complex quantum states~\cite{tan2025measuringnonabelianquantumphase} to Variational Quantum Algorithms, a more general class of hybrid classical-quantum algorithms that typically use classical optimisation or post-measurement selection to solve problems, with the Variational Quantum Eigensolver specifically being used to great effect in quantum chemistry ~\cite{Tilly_2022}.
In general, the logic of variational quantum algorithms, specifically of variational quantum optimisation algorithms (including but not limited to QAOA, ma-QAOA, NV-QWOA), move through several steps, going from an initial zero state of  $\ket{00...0}$, to an initialised state $\ket{\psi_{1}}$ which encodes the problem instance specific information, but where no processing has been done to manipulate the essentially still randomised output, to a final state $\ket{\psi_{2}}$ where, through applying unitary operators to the system, the probability distribution of the solution states has been altered to amplify the optimal and near-optimal states and suppress the undesirable non-optimal states. 
From here, many shots are fired so as to approximate the probability distribution of the final state with confidence, from which the algorithm uses classical methods to take this measurement data and calculate an objective function (such as the cvar or expectation value of the final state), to vary and tune hyperparameters to minimise the objective function, these overall circuits are often referred to as Variational Quantum Circuits (VQCs).

QUBO (Quadratic Unconstrained Binary Optimisation) is a NP-Hard combinatorial optimisation problem which has been previously used in attempting to solve the Closest String Problem on quantum annealing hardware~\cite{QUBO}, by rewriting in in a QUBO encoding.
To work around constraints in the types of problem that would be compatible with a QUBO representation, penalties can be introduced to the function, which mark invalid solutions by increasing their cost (assuming the minimisation of $f(x)$). One penalty might be for working in a subspace of the Hilbert space of a solution register, where solutions that lie outside the subspace but within the larger Hilbert space available in the system are considered invalid, and can't be presented as optimal solutions, for example, if an alphabet consists of the characters $1,2,3$, a QUBO implementation creates a Hilbert space that allows the character to be $1,2,3,4$, but all solutions containing a 4 are inherently not valid potential solutions.
The base version of QAOA (the Quantum Approximate Optimisation Algorithm), is a hybrid quantum-classical approximate algorithm used to solve combinatorial optimisation problems~\cite{QAOA}. It operates by iteratively applying a phase separation and a mixing Hamiltonian to an initial superposition, where the phase separator applies a local phase to states proportional to their cost, then the mixer applies a rotation of the state from the superposition proportional to its phase. Together this amplifies states with some proportionality to their quality. These are applied for a chosen number of iterations then the system is measured. This is then run for a number of shots to sample the probability distribution.
The phase separation unitary is generally defined as:
\begin{equation} 
\hat{U}_{phase}=e^{-i\gamma_k\hat{H}_{phase}},
 \end{equation} 
and the mixer unitary as:
\begin{equation} 
\hat{U}_{mixer}=e^{-i\beta_k\hat{H}_{mixer}},
 \end{equation} 
where $k$ is the iteration number and refers to the $k^{th}$ element in each parameter vector (indexed from zero), with $0\le k\le p-1$, where $p$ is the total number of iterations applied and the length of the parameter vectors, $\hat{H}_{phase}$ is the the Hamiltonian which encodes the cost function and $\hat{H}_{mixer}$ is the Hamiltonian which encodes the mixing.
The unitary for the QAOA can therefore be written in full as:
\begin{equation} 
\hat{U}_{QAOA}(\vec{\gamma},\vec{\beta},p)= \prod^{p-1}_{k=0}e^{-i\beta_k\hat{H}_{mixer}}e^{-i\gamma_k\hat{H}_{phase}},
 \end{equation} 
where the hybrid aspect of the algorithm comes in with the parametrisation of the two Hamiltonians, as each takes a parameter vector as input to scale their effects. After obtaining the probability distribution, it's used to calculate some measure of quality from the distribution such as expectation value or a cvar, which is used to then classically optimise the input parameter vectors.
There are many variants and algorithms which have evolved out from the original base QAOA, from ma-QAOA which adds more parameters to each layers to optimise over, QAOA+ which adds a problem independent layer of parameters to optimise over, to the QWOA which implements the mixer as a quantum walk over a graph to better preserve inherent problem structure~\cite{QAOA}.
The foundation of state vector simulation of a quantum computer is that for $N$ qubits (which would each be physically encoded as a two-level quantum system in a quantum computer), a classical computer needs to store $2^{N}$ complex state vectors containing the amplitude and complex phase of each state in the superposition. Gates are translated to unitary matrices, then applied in matrix form to these complex state-vectors through matrix multiplication, and the probability of observing each state can be taken from their final amplitudes "squared" (multiplied by their complex conjugate)~\cite{herculean}.
There are a wide variety of methods to do this including using tensor networks and the state-vector models. We primarily used state vector simulation of quantum circuit through Qiskit, mimicking how code would have to be written for a fault tolerant quantum computer and initially also using matrix exponentiation methods with classically pre-calculated cost functions and adjacency matrices used to implement the phase shift and mixer with its quantum walk as one exponentiated matrix each.

\subsection{Prior Attempts to Solve the Problem}

Given that these are NP-Hard problems, there are, by definition, no known classical exact optimisation algorithms with polynomial runtime or better.
There are, however, approximate and parameterised algorithms which can reach polynomial run times, at the cost of either not necessarily finding the optimal solutions, or by preemptively restricting the solution space.
For approximate algorithms, the user specifies a range of how far a solution can vary from the optimal solution, and the algorithm will find a solution in that range (potentially in polynomial time depending on the algorithm). Generally, lower-cost variance allowed from the best solution correlates to a longer runtime. As such, approximate algorithms are used when above-average quality solutions are sufficient as opposed to requiring the most optimal solution.
Parameterised algorithms constrain certain parameters to achieve a potential speed-up, by restricting the size of the solution space that is searched. The current fastest classical parameterised algorithm for the decision problem version of the Closest Substring Problem , runs with time complexity $\mathcal{O}(mn + md*61.86^{d})$~\cite{CSP} (with \textit{m} reference strings, substring length of \textit{n} and a maximum Hamming Distance of \textit{d} that a solution can have to any substring).
While this works well for small values of \textit{d}, real-world examples with significant numbers of mutations or errors in the reference string set would still run into exponential time scaling with $d$.
In addition, without knowing \textit{d} beforehand, a user would have to run progressively wider sweeps over increasingly large \textit{d} values to confirm if a solution existed for them (remembering that each of these sweeps is an NP problem itself), which may end up being less efficient than an approximate algorithm for many problems.

Dissanayake~\cite{QUBO} used a QUBO reformulation 
to attempt to solve the Closest String Problem, and while they have shown promising results, during the reformulation the objective function is actually changed to that of the Consensus Error String Problem, making direct comparison to this research incompatible, as it solves for an entirely different objective function. 
Due to the nature of the Consensus Error String Problem's objective function, Dissanayake can split the problem into subproblems, solving for each character separately, as the objective function takes sum of differences between each character in a reference string and the solution string, and then a sum of those Hamming Distances for each reference string. As such, for a problem instance with $m$ reference strings, each $n$ characters long, he can treat it as $n$ subproblems using the original Closest String objective function in which there are $m$ reference strings and $n=1$, solving for each character in the optimal solution separately and recombining them afterwards. This clever approach removes a string length dependence from the number of qubits required and substitutes it by multiplying the runtime by the number of characters instead. 
No benchmarking was done beyond four proof of concept tests using four different sets of reference strings, but there was a stated established maximum limit of 182 reference strings for their architecture, with a theoretical qubit scaling of $mn$ qubits being required on a fully connected quantum system, with harsher requirements being implied with limited system connectivity.

Unfortunately this approach is not compatible with the original Closest String Problem, as the max function requires comparison between each complete reference string and the complete solution string before minimisation, making this decomposition non-viable for the Closest String Problem's main formulation. 
In addition, doing these comparisons required for the max function in a gate-based system uses ancilla registers to store the results, whereas in quantum annealing this max function would have to be encoded into the Hamiltonian in terms of a QUBO problem, but this doesn't work as a max function is not quadratic or continuous. As such there would need to be multiple penalty terms introduced to attempt to enforce a max, and the number of penalties to enforce $dist_{max}\ge dist_1,dist_{max}\ge dist_2...$, would grow continuously with the number of reference strings, as each comparison would introduce a new condition. 
In addition, the incompatibility between using a non-binary alphabet with QUBO's objective function will create more added penalties, to mark invalid solutions created by the binary encoding when the alphabet size is not a power of two, as all unused letters will need penalties to mark solutions containing them, beyond the ones already used by Dissanayake, and those that a maximum function would require.
Even after this reworking, QUBO presents an inefficient spatial implementation in terms of the number of qubits required per reference string compared to the theoretical limit~\cite{QUBO}.

\subsection{The Non-Variational Quantum Walk Optimisation Algorithm}

The Non-Variational Quantum Walk Optimisation Algorithm (NV-QWOA) is a quantum approximate optimisation algorithm that solves combinatorial optimisations problems, designed by Tavis Bennett, as a descendent of the standard QAOA, with several key modifications and conditions added. This results in better exploitation of inherent problem structure and a significant reduction of the parameters, going from the two parameter vectors associated with standard QAOA to just three hyperparameters~\cite{tavis}. In addition, the three parameters in NV-QWOA are often problem type specific, having their optimal values show small variation between problem instances of similar size and structure.
The NV-QWOA has been tested and verified in applicability on the weighted maxcut, k-means clustering, maximum independent set and capacitated facility problems among others~\cite{bennett2024analysisnonvariationalquantumwalkbased}.
In the NV-QWOA, rather than setting independent parameters for each value of $t_j$ and $\gamma_j$, where $0\le j < p$, one value for $\gamma$ and $t$ are each accepted as input, and the values of $t_j$ and $\gamma_j$ are defined such that:
\begin{equation} 
\gamma_j = \gamma (\beta+j\frac{1-\beta}{p-1}),
 \end{equation} 
\begin{equation} 
t_j = t (1-j\frac{1-\beta}{p-1}),
 \end{equation} 
such that $\gamma_j$ increases linearly over the domain $[\beta\gamma,\gamma]$ and $t_j$ decreases linearly over the domain $[\beta t,t]$.
The NV-QWOA follows these steps:
\begin{enumerate}
\item The cost unitary is applied to the circuit to calculate the cost of each solution state (in superposition).
\item A phase shift proportional to the cost of each solution state is applied to the solution register, controlled off the calculated costs in the ancilla register which is scaled by the hyperparameters $\gamma$ and $\beta$, along with $\sigma$, the standard deviation of the cost of the output state and $\alpha=\pm1$, which is negative when minimising the cost function and positive when maximising (such that we set $\alpha=-1$ in our research). This phase unitary is represented in Figure 1, 
\begin{equation} 
U_{P}(\gamma_j) = e ^{-i\beta \operatorname{Q}(\alpha\frac{\gamma_j}{\sigma}) },
 \end{equation} 
where $Q$ is the diagonal $N*N$ matrix (where $N=k^n$, the size of the solution space) containing the ordered eigenvalues of each solution string to the cost operator.

\item A continuous time quantum walk (CTQW) over the mixing graph $A$ for time $t$ is applied to the input register. 
This mixing unitary is defined as:
\begin{equation} 
U_{M}(t_j) = e ^{-i t_j A},
 \end{equation} 
where $A$ is the mixing graph (or adjacency matrix) which connects neighboring solution states for the CTQW. In the case of the Closest String and Substring Problems, solutions separated by a Hamming distance of 1 from each other are considered to be neighbors. ~\cite{asensio2025sensitivityhamminggraphs}
\item Steps 1-3 are repeated for a total of $p$ iterations, to amplify the probability of measuring the optimal or near optimal solution states, with each iteration further amplifying solutions based on their quality (minimising their cost), this will eventually begin to plateau at an instance-specific number of iterations. In some cases, excessive iterations can erode the amplification of optimal states, making the choice of $p$ important to effective implementation of the algorithm. The implementation of the paired phase and mixing unitaries for $p$ iterations is formally defined as:
\begin{equation} 
U_{NV-QWOA}(\beta ,\gamma, t,p) = \prod^{p-1}_{j=0} U_M (t_j) U_Q(\alpha\frac{\gamma_j}{\sigma}).
 \end{equation} 
\begin{figure}[htb]
    \centering
    \includegraphics[width=0.3\linewidth]{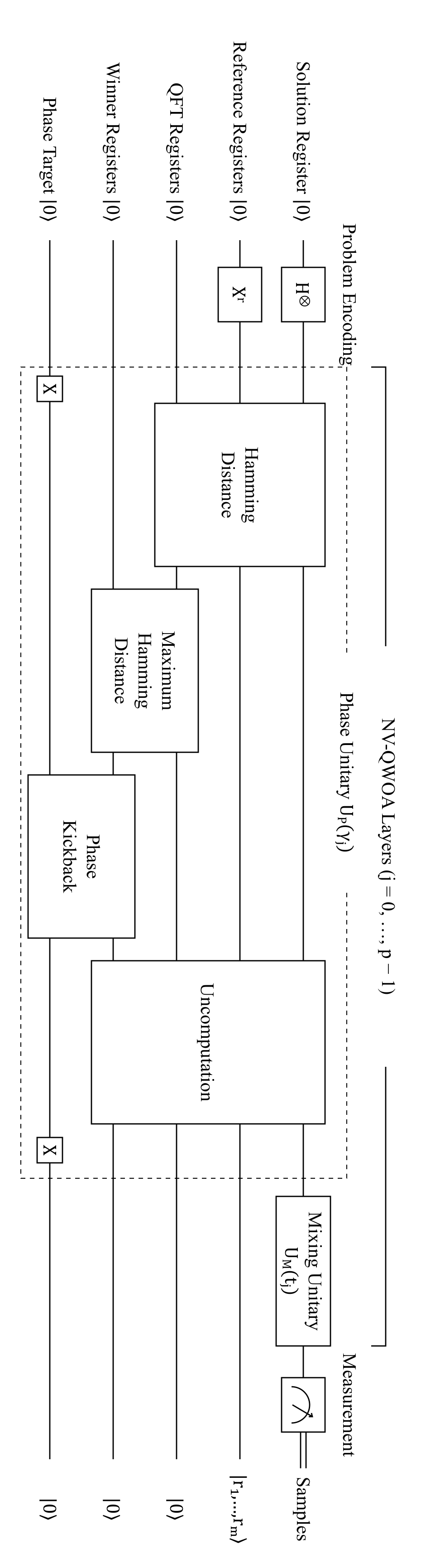}
    \caption{The full generalised circuit implementation of the NV-QWOA for the Closest String Problem, implementing the problem encoding, steps 1-4 as described, then measurement.}
    \label{fig:placeholder}
\end{figure}
\end{enumerate}
A solution's subshell value is the smallest Hamming distance from it to any of the optimal solutions (the solution or solutions with the lowest cost for an instance make up the optimal solution set). The set of states in a subshell are the collection of nearest neighbors to the optimal solution set within a set hamming distance ( a distance of two for subshell two, etc.). 
It can be used as a representative of solution quality when calculating means between instances, where one instance may have optimal solutions with a cost of 0 and another instance may have optimal solutions with a cost of 5, which isn't a visually useful way of representing optimal solution amplification, whereas both of those solutions are in subshell 0 of their respective instances.

\section{Quantum Circuit Implementation}\label{Quantum Circuit Implementation}

Qiskit is IBM's quantum computing Python library for creating, transpiling, classically simulating quantum circuits and running them on quantum hardware.  Qiskit-Aer is the software package responsible for the classical ideal and noisy simulation of these circuits~\cite{qiskit2024}. 
We made of use of Qiskit's QuantumCircuit class, evolved it with the problem instance and NV-QWOA ansatz, simulating the state through AerSimulator's state-vector simulation for 1024 shots.
Qiskit uses a little-endian significance ordering convention. This means that by default, if you had a circuit diagram with three qubits, $q_0,q_1,q_2$, where $q_0$ is the top wire in the diagram, $q_1$ is in the middle and $q_2$ is the bottom wire in the diagram, the state vector representation of this that Qiskit gives would be $\ket{q_2q_1q_0}$, and it is implicit that $q_2$ is the most significant bit and that $q_0$ is the least significant bit. We followed this ordering in my circuit within each subregister, and it is implicitly used in any circuit diagrams provided in this research.

\subsection{Phase Unitary Implementation}

Implementing the phase unitary occurs through first calculating the Hamming Distance between the solution register and each reference subregister, comparing and writing forward the larger quantum state (in the case that the two Hamming Distances are equal it is irrelevant which value is written forward), applying a phase shift to the solution register via phase kickback proportional to the largest Hamming distance of these values, then uncomputing all of steps prior to the final phase shift, to disentangle the ancilla and solution registers and prevent this entanglement from impacting the mixing.
The uncomputing is done by applying the Hermitian conjugate of the overall cost unitary. In practice this means the Hermitian conjugate of every individual gate applied in the cost unitary is applied, in reversed order to when they were first applied to calculate the cost.

The Hamming distance is calculated in two parts. We first determine if each individual character in the reference string is different in value to the equivalent character in the solution superposition, then use Quantum Fourier Addition (also known as a Draper Adder) to add up the sum of character differences, outputting the Hamming distance encoded in binary.
The implementation of the Hamming distance calculation is dependent on the value of $k$ in the problem instance. $k<0$ is a meaningless and therefore invalid set of values (defining the alphabet to have a negative number of characters is meaningless), $k=0$ describes an empty set, which is useless in the context of this problem and thus also deemed invalid, and $k=1$ describes problem instances in which all reference and solution strings are just $n$ repetitions of the same character (which means that the number of potential solution strings is $k^n=1^n=1$), and thus trivial to solve, leaving only the domain of $k\in \mathbb{Z}, k\ge2$ to be considered for implementation, as only here do problems become non-trivial and meaningful.

In the case where $k=2$, calculating whether a character in the solution superposition and a reference string are different simply requires applying a controlled Pauli-X gate (CNOT gate), controlled on the qubit encoding that character in the solution superposition and targeting the qubit encoding the character of corresponding position within the reference string. This operation leaves the target qubit in the reference register in the $\ket{0}$ state if the control and target qubits were initially together in the $\ket{00}$ or $\ket{11}$ states and in the $\ket{1}$ state if they were initially in the $\ket{01}$ or $\ket{10}$ states.
When $k>2$, multiple qubits are required to encode the potential values of one character. The above process is applied to each corresponding qubit in the solution register and reference subregister. When a character differs from the solution to reference string, one or more of the difference target qubits must be in the $\ket{1}$ state, as the binary encoding of the two characters must differ in at least one position. As such we needed a way to implement the quantum equivalent of an (inclusive) OR gate that takes all difference qubits as inputs and outputs to an ancilla. 
The method we used was to create an ancilla register with $mn$ qubits, where each reference string character is assigned one additional ancilla qubit from this character register, and has a CNOT controlled by each difference qubit corresponding to that character act on it. 

From here, it depends on the value of $\lceil \log_2( k)\rceil$, which is the number of qubits required to encode one character. If this value is equal to 2, then in addition to each difference qubit controlling a CNOT on the character qubit, every combination of the pairs of two of the difference qubits act as the controls to apply a CCNOT (double controlled Pauli-X gate) to the character qubit. 
\begin{figure}[H]
    \centering
    \includegraphics[width=0.7\linewidth]{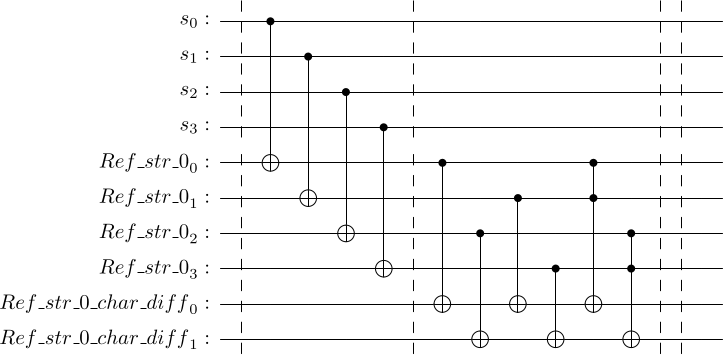}
    \caption{The section of the Hamming Distance circuit that marks whether each character in the solution and reference registers differ.}
    \label{fig:placeholder}
\end{figure}
When the value of $\lceil \log_2( k)\rceil=2$ (as shown in Fig. 1, where $k=4$), there is only one pair which does this, meaning cases where $\lceil \log_2( k)\rceil=2$ will have a total 2 CNOT gates and one CCNOT gate acting on each character qubit.
As we go to $\lceil \log_2( k)\rceil=3$, there will be a total of 3 CNOT gates, 3 CCNOT gates, and a new layer which applies every permutation of a CCCNOT gate (there is one in this case) to the target qubit.
This general trend continues, such that, for $\lceil \log_2( k)\rceil=a$, each combination of qubits in the reference subregister corresponding to the appropriate character, where the combination contains $a$ or less qubits, will act as the controls for a multi-controlled NOT gate, targeting the character qubit. 
In computing the Hamming distance, we implemented Quantum Fourier Addition to sum the values of the character qubits for each reference string into the QFT subregister corresponding to that reference string. This results in the Hamming distance between the solution superposition and each character string being encoded in binary in the QFT register at the end of this process. 

Quantum Fourier Addition is an efficient tool for quantum arithmetic, which computes sums in the Fourier (frequency) basis by using the Quantum Fourier Transform (QFT) and inverse QFT to transform between the computational and frequency bases~\cite{draper2000additionquantumcomputer}.
The standard QFT gate implementation makes use of SWAP gates, but in the interest of gate efficiency, we implemented a swapless version. This can effect management of endianness while working in the frequency basis, but cancels out after applying the swapless inverse QFT (the Hermitian conjugate of the QFT unitary). 
After the initial QFT gate, conditioned phase shifts weighted by the binary value of the control qubits are applied to the QFT register, such that when we apply the inverse QFT to the register afterwards, the result is the binary encoded sum of the value of the control qubits. The endian order is reversed in this gate due to the swapless QFT applied previously.

If $k=2$, the controls of these conditioned phase shifts are in their respective subregister of the reference register (as in fig), whereas for $k>2$, the controls come from the character register. 
This implementation however, scales poorly in circuit depth and size, and after realising that the difference counts were functionally implementing a quantum equivalent to an inclusive OR gate, we found that IBM had a far more efficient implementation~\cite{IBM_OR} for the OR gate available.
We combined their OR gate with the already implemented QFA such that we didn't require the additional ancilla needed for the original design, such that when the OR would originally have outputted 1, it now was the integrated control for the phase rotation of that difference in the QFA, such that the total Hamming distance calculation is done as such, (not including the QFT basis change):

\begin{figure}[H]
    \centering
    \includegraphics[width=0.9\linewidth]{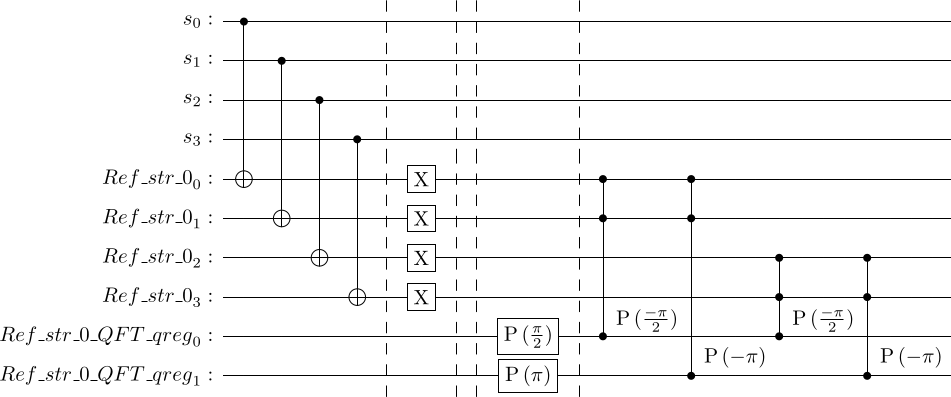}
    \caption{Shown is the implemented Hamming distance circuit, with QFT and inverse-QFT removed for brevity, performing Quantum Fourier Addition combined with an OR Gate, such that if a character represented by more than one qubit differs it is still encoded.}
\end{figure}

\begin{figure}[H]
    \centering
    \includegraphics[width=1\linewidth]{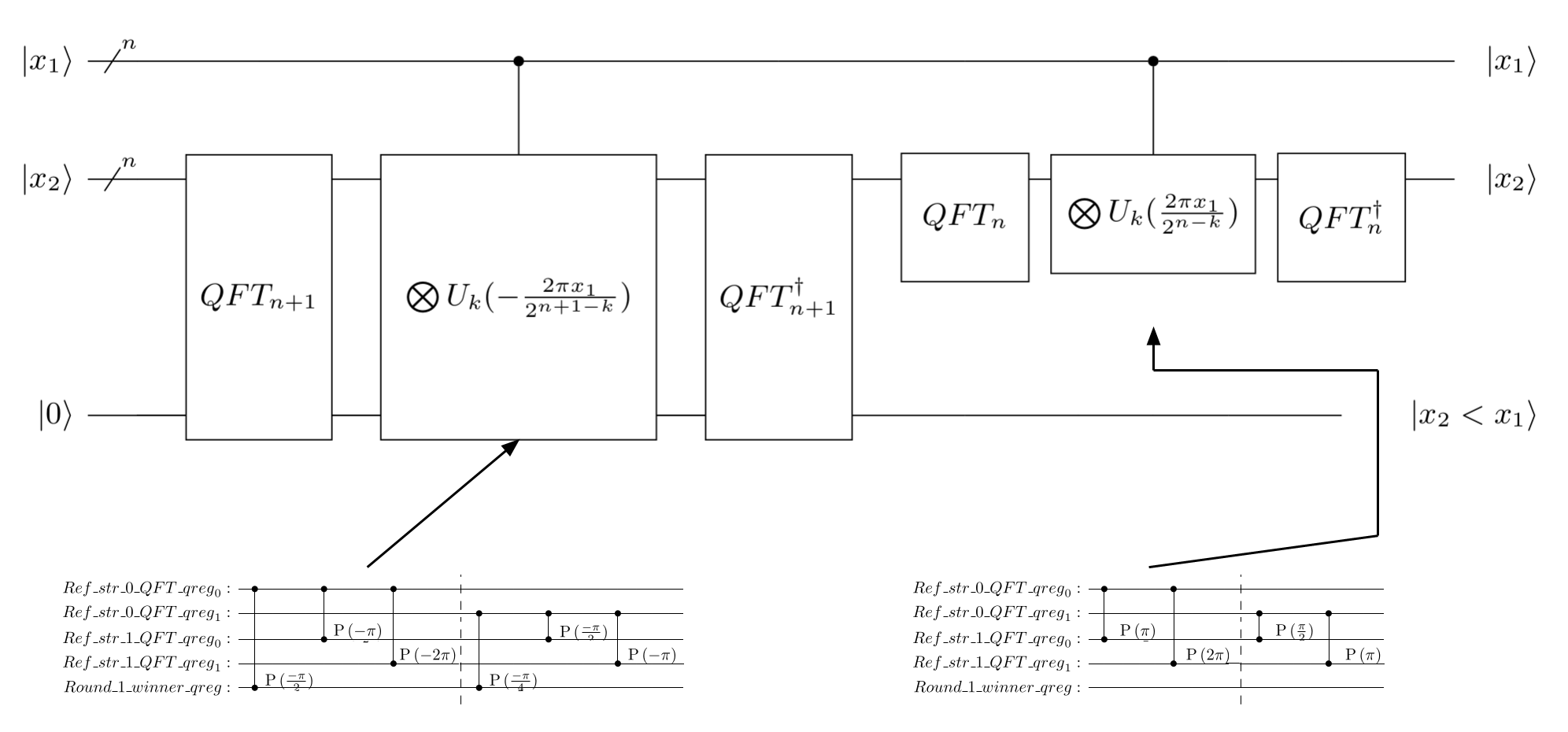}
    \caption{The Quantum State Comparator (main circuit diagram), as designed by Yuan et al.~\cite{statecomparator} is the top of the figure with specific implemented circuits of the $U_k$ gates shown below. In the bottom left, an example of the implemented first half of the state comparator is shown, (which performs the comparison $x_1>x_2$) by applying the $\operatorname{QFT}$ unitary, applying phase shift $U_{j}(-\frac{2\pi x_1}{\lceil\log_2(n\rceil+1-j)})$ (relabelled for clarity) in the bit-order reversed frequency bases iteratively from the LSB ($j=0$) through the register and the added comparison qubit, then applying the $\operatorname{QFT^{\dagger}}$ unitary to return to the computational basis. In the bottom right, an example of our implemented second half of the state comparator is shown, performing the second set of basis transformations and phase rotations then mapping the previous state to the desired state of $\ket{x_1,x_2,x_1>x_2}$. }
    \label{fig:placeholder}
\end{figure}
We made use of the QFT based quantum state comparator detailed by Yuan et al. in Fig. 4, which takes as input the two quantum states $\ket{x_1}$, $\ket{x_2}$ along with an ancilla qubit $\ket{0}$, and returns $\ket{x_1}$, $\ket{x_2}$ and $\ket{x_1>x_2}$. 
It makes use of the QFT, with the authors providing the required modifications to the phase rotation components required when using the swapless QFT variant (implemented as earlier described) . The overall circuit for the state comparator as detailed in their paper is shown in Fig. 4, where $x_1$ and $x_2$ are the values of the two quantum states $\ket{x_1}, \ket{x_2}$, $n$ in this figure refers to the number of qubits required to encode one of $\ket{x_1}$ (or $\ket{x_1}$, as this comparator assumes the two registers are of equal length and use binary encoding instead of one-hot) and $k$ is the $k^\text{th}$ target qubit, indexing from the least significant bit as $k=0$ through to and past the most significant bit to the final, initially $\ket{0}$ qubit. 
In our case, there are $\lceil \log_2(n)\rceil$ qubits required to encode$\ket{x_1}$ as we make comparisons between the QFT registers.

The first $\operatorname{QFT^{\dagger}}$, phase rotation and $\operatorname{QFT^{\dagger}}$ group act to perform the comparison of states between $\ket{x_1}$ and $\ket{x_1}$, where N is the period of this Fourier basis, performing the mapping: 
\begin{equation} 
\ket{x_1,x_2,0}\rightarrow\ket{x_1,N+x_2-x_1,1}_{x_1>x_2}+\ket{x_1,x_2-x_1,1}_{x_2\ge x_1} 
\end{equation} 
Despite the literature quite extensively detailing implementations of quantum state comparators, there appeared to be an absence of an explicit desciption of the implementation of an exact maximum or minimum function. As such we designed two circuit architectures (using the CCNOT and CSwap gates respectively) to implement them. 
The swapless version requires $\lceil\log_2(n)\rceil$ additional ancilla qubits for every comparison after the first (for $(m-1)\lceil\log_2(n)\rceil$ qubits added in total). 
For the i$^{\text{th}}$ qubit in state $\ket{x_1}$ it uses that qubit and the output of the state comparison $\ket{x_1>x_2}$ the controls of a CCNOT gate, which targets the i$^{\text{th}}$ in the ancilla subregister added for that comparison, then applies a Pauli-X gate to the output of the state comparison $\ket{x_1>x_2}$ and repeats this process with each qubit in state $\ket{x_1}$, which copies the first state to the new subregister if $x_1>x_2$ and the second if $x_1\le x_2$. This causes no phase-kickback as the new register is initialised in the zero state. This process of comparison between the winner of the previous comparison and the next reference string's Hamming distance superposition's is repeated until all the final winner of the last comparison has been compared to every other Hamming distance superposition. This is thus the max of the list of Hamming distances between the solution superposition and all reference strings in the problem instance encoded in the circuit.  

In the swap version, rather than copy the state forward, the state comparison $\ket{x_1>x_2}$ is used to control a series of swaps between the pair of qubits of the same bit significance in $\ket{x_1},\ket{x_2}$, swapping the states on the register, such that the state encoding a larger value is now on the lower register. 
In both cases, this process of comparison between the winner of the previous comparison and the next reference string's Hamming distance superposition's is repeated until all the final winner of the last comparison has been compared to every other Hamming distance superposition. This is thus a deterministic maximum of the list of Hamming distances between the solution superposition and all reference strings in the problem instance encoded in the circuit. The difference being that after each comparison, one method writes the winner forward onto new ancilla and the other shuffles it down through existing ancilla.

\subsection{Deterministically Choosing $\gamma$}
In the original NV-QWOA, it is recommended to start at $\gamma=\frac{1}{\sigma}$, where $\sigma$ is the standard deviation in cost of the initial state, and this $\sigma$ is meant to constrain the phase separation between states such that it is roughly bounded within $[1,e^{i\pi}]$.
However at smaller scales, we found examples where this model starts to break down. Specifically, one such instance was an $n=4$ instance with $\sigma\approx 0.6$, such that the phase rotation on states with a cost of 4 was $\approx e^{2.1\pi i}=e^{0.1\pi i}$, causing a significantly decrease in the phase separation between the highest and lowest cost states, which should be maximally separated in the ideal case.
We thus investigated how defining $\gamma$ deterministically based on the problem instance parameters, could be used to avoid scaling it by the standard deviation of cost, which required heuristic estimates. As this standard deviation scaling was in place to prevent causing a phase shift greater than $e^{i2\pi}$ causing the phase to become degenerate between edge case states, we implemented a deterministic gamma, such that: $\gamma=\frac{\pi}{2(n+1)}$, as this constrains gamma by the maximum cost a state in a problem instance of size $n$ can have ($n$). This applies a phase shift of $e^{\frac{\pi i}{2}\frac{Cost}{n+1}}$ which is always in the domain $[e^{0},e^{\frac{\pi}{2}i})$, choosing $n+1$ instead of $n$ to remove degeneracy of states with maximum and zero cost. 
This caused significant improvement in the amplification of optimal solutions (without requiring the optimisation of $\gamma$, making the algorithm more non-variational in accordance with its name).
This speeds up the optimisation process by an order, reducing it to 2 parameters to optimise, or 1 if setting $\beta=\frac{1}{p}$, although even the unoptimised results produced a mean probability greater than $80\%$ of observing the optimal solution subset for each Closest String problem size investigated.
In addition, it is clear from the results that using this fixed gamma with or without the initial cost standard deviation scaling produces almost identical results, with the unscaled results being slightly better as problem size increases, as the fixed gamma implements the maximum phase separation which still prevents phase wrapping around $2\pi$, and decreasing it by this scaling factor is therefore reducing phase separation and thus relative amplification for no benefit.

\subsection{Mixing Unitary}
The mixing unitary performs a continuous time quantum walk over the mixing graph~\cite{qwalks}. The mixing unitary acts on the solution superposition state such that it redistributes amplitude between states connected on the graph. This combined with the phase separation causes amplitude to build over iterations on the states with small positive phases and decrease on states with large positive phases (flipping the sign changes the algorithm to maximising the cost), amplifying solutions according to cost. 
The mixing graph for the Closest String Problem (as we define the cost function), is the graph where all solutions are labelled vertices, and solutions differing in one character (separated by a Hamming distance of 1) are connected (have an edge between them). These set of graphs are known as Hamming graphs, of which the hypercube is the special case of a Hamming graph with $k=2$~\cite{Harary1988}. In NV-QWOA, the mixers using a Hamming graph are referred to as integer mixers.
Using the implementation of the integer mixer described by Bennett~\cite{bennett2024analysisnonvariationalquantumwalkbased}, the solution register has the custom unitary $e^{-itA}$, applied to it, using Qiskit's UnitaryGate class ~\cite{IBM_Unitary}, which performs the CTQW over the relevant Hamming graph (which is the hypercube for $k=2$).

\subsection{Space Complexity and Circuit Layout}

There were seven quantum registers and a classical register which the final measurement reads out to. Of the seven quantum registers, there were the solution register and six ancilla registers. All quantum registers are initialised in the zero state. 
The solution register then has a Hadamard gate applied to each qubit to create the state of equal superposition in this register, which is the Hilbert space which we will refer to as the solution space (which will be the same for all problem instances with the same values for $n$, $m$ and $k$), where $n$ is the length of the strings, $m$ is the number of reference strings in the problem instance and $k$ is the size of the alphabet.
The solution register's size, in terms of qubits required, scales according to $q_{sol}=n\lceil log_{2}(k)\rceil$. The classical register will contain an equal number of classical bits to qubits in the solution register, as the solution register is read out to it.
For indexing reasons, when designing the circuit we chose to use six ancilla registers rather than one, so we could more easily iterate over and between different sections of the circuit when implementing multi-qubit gates. 
These six ancilla registers are, the reference register, the character register, the QFT register, the winner register, the winner slot register and the phase target register.

The reference register is used to encode the reference strings. It is designed such that that problem instance inputted is formatted such that the alphabet of characters is assigned uniquely to integers, starting from zero and increasing by steps of 1 (an alphabet of $\{a,b,c\}$, would be mapped to $\{0,1,2\}$, etc.). We used a little-endian binary encoding, to convert the problem instance from a Numpy array of the above form, where each row is one reference string and each array element is a character, to a flattened vector.
This vector is directly read as the controls to Pauli-X gates in the reference register, with each character taking the value of 1 or 0, therefore encoding reference strings in binary in the circuit. 
The reference register's size is $q_{ref}=mn\lceil log_{2}(k)\rceil$, with each reference string requiring $n\lceil log_{2}(k)\rceil$ qubits.
The character register is used as an intermediary to calculate whether characters are different when the alphabet is larger than $k=2$, and is not necessary for $k=2$. When required, it is of size $q_{char}=mn$ with each reference string requiring $n$ qubits, one per character in the reference string to mark if that character differs from the solution superposition's respective character.

The QFT register is to encode the values of the Hamming distances between the solution superposition and the reference strings, and is split into subregisters encoding each Hamming distance. It is labelled the QFT register as the Hamming distances are summed from the difference counts using QFT addition, which leaves the Hamming distances encoded in binary. As there are $2^n+1$ possible costs that could be encoded ($[0,2^n]$) this register is of size $q_{QFT}=m\lceil log_2(n+1)\rceil$, with each subregister being of size $\lceil log_2(n+1)\rceil$.

The winner slot register is used to store the results of comparisons by writing the greater quantum state forward. It is used in the original swapless version of the circuit, and can be removed for a more space-efficient version of the circuit by using controlled swap gates, which shuffle the largest quantum state down the circuit rather than continuously writing the winner forward. When implemented it is of size $q_{win slot}=(m-1)\lceil log_2(n+1)\rceil$.
The winner register is single ancilla qubit required for each comparison to store each value $x_1\ge x_2$. It is of size $q_{winner}=m-1$.
The final register is the phase target register, which only requires one qubit for any circuit, to which the phase shift controlled from the calculated cost superposition is applied, to cause phase kickback proportional to the cost to the solution register. It is of size $q_{target}=1$.

As such, depending on the problem and circuit version, the total circuit size is 
\begin{equation} q_{circ}=n(m+1)\lceil \log_{2}(k)\rceil+m\lceil \log_2(n+1)\rceil+m+q_{win slot}+q_{char}.
 \end{equation} 
For the swapless, $k=2$ case: 
\begin{equation} 
q_{circ}=n(m+1)+(2m-1)\lceil \log_2(n+1)\rceil+m.
 \end{equation} 
For the swapless, $k>2$ case: 
\begin{equation} 
q_{circ}=n(m+1)\lceil \log_{2}(k)\rceil+(2m-1)\lceil \log_2(n+1)\rceil+m(n+1).
 \end{equation} 
For the controlled swap, $k=2$ case:
\begin{equation} 
q_{circ}=n(m+1)+m\lceil \log_2(n+1)\rceil+m.
 \end{equation} 
Finally, for the controlled swap, $k>2$ case:
\begin{equation} 
q_{circ}=n(m+1)\lceil \log_{2}(k)\rceil+m\lceil \log_2(n+1)\rceil+m(n+1).
 \end{equation} 
As such, for all cases, the circuit size scales to:
\begin{equation} 
\mathcal{O}(nm \log(k)).
 \end{equation} 
Achieving a space complexity scaling that is linear with $n$, linear with $m$ and logarithmic with $k$.
We were able to further optimise this spatial scaling at the last minute, through finding a more efficient method of implementing the OR gate which requires only one ancilla per OR gate. ~\cite{IBM_OR} By modifying the OR gate, we could implement the Hamming distance calculations without adding any additional ancilla for any value of $k$, as shown in Figure 3.4. This reduced both the controlled swap and swapless implementations for $k>2$ circuits' final qubit count by $mn$ qubits.
As we implemented the controlled swap version to generate that, the final implemented qubit counts for the Closest String problem circuit were left as:
\begin{equation} 
q_{circ}=n(m+1)\lceil\log_2(k)\rceil+m\lceil \log_2(n+1)\rceil+m.
 \end{equation} 
\subsection{Time Complexity and Circuit Depth}
In calculating the order of the time complexity, only terms that scale with problem instance parameters ($n,m,k,p$) are relevant, with other constant terms not effecting the scaling. One layer of gates can be considered as the collective unitary which describes each qubit having one gate acting on (or being controlled by) it, where empty spaces are replaced by the identity gate. The time-complexity (in quantum computing) of a circuit describes how the number of layers required scales with the problem instance parameters.
Encoding the reference strings only requires one Pauli-X gate per qubit, similarly, creating the initial superposition of solution states only requires one Hadamard gate per qubit, giving a total of one layer being required to initialise the circuit (these two operations are simultaeneous).
In the phase separator gate, the character difference operations require one layer of CNOT gates, with each qubit involved only being used once, adding one layer per iteration of the algorithm, for a total of 1 layer per iteration.
If $k>2$, there will be $\lceil \log_2( k)\rceil$ qubits encoding a character difference, with $\binom{\lceil \log_2( k)\rceil}{1}$ CNOT gates required, $\binom{\lceil \log_2( k)\rceil}{2}$, CCNOT gates required, etc., and given each gate acts on the same qubit, they can be treated as individual layers, giving a total layer contribution per iteration of ($L_{combo}$) of:
\begin{equation} 
L_{combo}=\sum^{\lceil \log_2( k)\rceil}_{c=1}\binom{\lceil \log_2( k)\rceil}{c}=\sum^{\lceil \log_2( k)\rceil}_{c=1}\frac{\lceil \log_2( k)\rceil!}{c!(\lceil \log_2( k)\rceil-c)!}.
 \end{equation} 
For the QFT gate, there are $\lceil \log_2(n+1)\rceil$ qubits in the register. Implementation requires for each qubit in the register there is one Hadamard gate applied ($\lceil \log_2(n+1)\rceil$ layers). Then there is a controlled phase gate applied between each qubit and all qubits lower than it in the circuit's subregister such that it produce the total number of layers per iteration:
\begin{equation}
L_{QFT}=\sum^{\lceil \log_2(n+1)\rceil-1}_{i=0}(\lceil \log_2(n+1)\rceil-i)=\frac{\lceil \log_2(n+1)\rceil^2+\lceil \log_2(n+1)\rceil}{2}.
 \end{equation} 
The Quantum Fourier addition's conditioned phase shifts iterate over $n$ and $\lceil \log_2(n+1)\rceil$ in a nested loop, requiring a maximum of $L_{QFA}=n\lceil \log_2(n+1)\rceil$ layers (without optimally sorting the operations to parallelise some gates).
The inverse QFT implemented requires the same number of gates as QFT, such that 
\begin{equation}L_{invQFT}=\frac{\lceil \log_2(n+1)\rceil^2+\lceil \log_2(n+1)\rceil}{2} \end{equation}  
per iteration.
As such the total Hamming distance evaluation requires 
\begin{equation} 
L_{hamming}=1+\lceil \log_2(n+1)\rceil^2+(n+1)\lceil \log_2(n+1)\rceil +L_{combo}.
 \end{equation} 
The comparison is implemented via a QFT gate, the phase rotations, an inverse QFT gate, then this is repeated on a different register with different phase rotation angles.
The first QFT gate acts on one QFT subregister and and additional ancilla qubit, thus it requires:
\begin{equation} 
L_{QFT2}=\frac{(\lceil \log_2(n+1)\rceil+1)^2+\lceil \log_2(n+1)\rceil+1}{2}.
 \end{equation} 
The following set of phase rotations are controlled by the first state and target the second state along with an ancilla qubit, giving:
\begin{equation} 
L_{rotate}=\lceil \log_2(n+1)\rceil(\lceil \log_2(n+1)\rceil+1).
 \end{equation} 
The first inverse QFT gate will thus require:
\begin{equation} 
L_{invQFT2}=\frac{(\lceil \log_2(n+1)\rceil+1)^2+\lceil \log_2(n+1)\rceil+1}{2}.
 \end{equation} 
The second QFT gate only acts on one QFT register, meaning it requires the same number of layers as the QFT gate found in the Hamming distance calculation, such that:
\begin{equation} 
L_{QFT3}=\frac{\lceil \log_2(n+1)\rceil^2+\lceil \log_2(n+1)\rceil}{2}.
 \end{equation} 
The second set of rotations are controlled on the first state, and target the second state, thus requiring:
\begin{equation} 
L_{rotate2}=\lceil \log_2(n+1)\rceil^2.
 \end{equation} 
As such, one comparison between strings requires:
\begin{equation} 
L_{comparison}=4\lceil \log_2(n+1)\rceil^2+5\lceil \log_2(n+1)\rceil+2.
 \end{equation} 
The larger state is then written forward (this would have the same layer requirement for the implementation using swaps instead of write-forward ancilla, replacing CCNOT gates with CSwap gates), requiring:
\begin{equation} 
L_{write}=2\lceil \log_2(n+1)\rceil+1
 \end{equation} 
layers, as the first state is conditionally written forward to the next ancilla subregister, the $\ket{x_1>x_2}$ qubit has a NOT gate applied to it, then the other state is conditionally written forward (such that only the larger state is written forward).
In one phase gate, there are $m-1$ comparisons made per iteration, making the total layers for the comparison layer equal to:
\begin{equation} 
L_{max}=(m-1)(4\lceil \log_2(n+1)\rceil^2+7\lceil \log_2(n+1)\rceil+3).
 \end{equation} 
The final step in the phase gate before uncomputing is applying the phase separation to the phase target register, where each qubit encoding the cost superposition applies a weighted conditional phase shift to the singular phase target qubit, such that:
\begin{equation} 
L_{shift} =  \lceil\log_2(n+1)\rceil.
 \end{equation} 
As such, including the uncomputation, implementing the phase gate for one iteration will require:
\begin{equation} 
L_{phase}=2L_{hamming}+2L_{max}+L_{shift},
 \end{equation} 
where $L_{combo}$ is zero for $\lceil\log_2(k)\rceil\le1$, and takes the value of the sum earlier defined for $\lceil\log_2(k)\rceil>1$.
Applying the CTQW over the integer mixing graph is split into applying phase rotations to each qubit in the solution register (this only requires two layers per iteration as the phase rotations are classically calculated and are applied at the beginning and end of the CTQW), applying a NOT gate to each qubit in the solution register (requiring two layers per iteration for the same reasons as in the phase rotations), and applying either a phase rotation, or multi-controlled phase rotation to the most significant bit representing each character in the solution register. These rotations require $n$ layers to implement per iteration, meaning the mixer requires:
\begin{equation} 
L_{mixer}=4+n.
 \end{equation} 
As such, the total layers (or circuit depth) required to implement $p$ iterations of the NV-QWOA through this circuit design is:
\begin{equation} 
L_{NV-QWOA}=1+p(L_{phase}+L_{mixer)}.
 \end{equation} 
Meaning that the time complexity of this implementation of the NV-QWOA is generally of order:
\begin{equation} 
\mathcal{O}(p(m\log(n)^2 
+L_{combo}
).
 \end{equation} 
When $k>2$, the time complexity of this implementation of the NV-QWOA is of order:
\begin{equation} 
\mathcal{O}(pm\log(n)^2
+
p\sum^{\lceil \log_2( k)\rceil}_{c=1}\frac{\log(k)!}{c!( \log(k)-c)!}
).
 \end{equation} 
Given:
\begin{equation} 
\mathcal{O}(\frac{\log( k)!}{c!(\log( k)-c)!)})=\mathcal{O}(\log(k)^{min(\log(k),\log(k)-c})).
 \end{equation} 
Since $c\ge1$:
\begin{equation} 
\mathcal{O}(\log(k)^{min(\log(k),\log(k)-c}))=\mathcal{O}(\log(k)^{\log(k)-c})).
 \end{equation} 
Giving us, when $k>2$. a time complexity of:

\begin{equation} 
\mathcal{O}(pm\log(n)^2
+
p\log(k)^{\log(k)-1}
).
 \end{equation} 
We were able to then optimise this gate scaling by implementing a more efficient implementation of the OR gate, used to find character differences in the Hammind distance calculation. 
The OR gate implementation from the Qiskit documentation~\cite{IBM_OR} requires one ancilla qubit. In our case it would implement an OR function on the difference qubits by applying one Pauli-X gate to each of the difference qubits and the ancilla qubit, then using the difference qubits as the controls to an MCX gate (multi-controlled X gate) which targets the ancilla. As the ancilla is in the $\ket{1}$ state, it stays ends in the $\ket{1}$ state if any to all of the difference qubits began in the $\ket{1}$ state as one of the controls is in a $\ket{0}$ state, and only flips if the difference qubits were in an all zero state. 
We modified this to work without needing additional ancilla in the circuit, as this output ancilla would have just been used as the control for a controlled phase gate in the quantum fourier addition. Instead, we changed the order of operations such that the QFT gate is applied to the QFT registers first, then uncontrolled phase gates are applied to the qubits in the QFT register, equivalent to the phase that would've been applied by a controlled phase controlled on the character register previously. From here a multicontrolled phase gate acts on each of the qubits, applying a negative phase equal to the phase just added, which cancels out the effects to a net zero phase shift if the OR gate outputs zero, and applies the required quantum fourier addition's phase shift if the OR operation gives 1. 
This takes 2 layers total for everything prior to the MCP gates, which then take $\lceil \log_2(n+1)\rceil$ layers total, as each reference subregister and QFT subregister can perform this simultaneously. 
This reduces $L_{combo}$ to $2p(\lceil \log_2(n+1)\rceil+2)$ layers total after uncomputing and $p$ iterations, bringing the total layer count down to:
$
3+n+6pm+p(2m-6)\lceil \log_2(n+1)\rceil^2+(14pm-10p+1+2pn)\lceil \log_2(n+1)\rceil \\
$ and the time complexity order to: 
\begin{equation} 
\mathcal{O}(pm\log(n)^2+pn\log(n), 
)
 \end{equation} 
with no $k$ dependence for circuit depth scaling.

For a fixed alphabet size $k$, the circuit depth scales as $\mathcal{O}!\left(pm\log^2 n+pn\log n\right)$, where $p$ is the number of NV-QWOA iterations, $m$ is the number of reference strings, and $n$ is the string length. This expression is linear in $p$, with a term linear in $m$ and polylogarithmic in $n$, together with a term proportional to $n\log n$. A fixed alphabet is relevant to many applications, including DNA sequences ($k=4$) and text restricted to the 26 English letters ($k=26$). These circuit resource requirements motivate further investigation of a potential speedup over known classical algorithms. Establishing an end-to-end quantum advantage requires evaluating the total computational cost rather than circuit depth alone. This includes determining how the NV-QWOA iteration count ($p$) scales with problem size, accounting for the measurement repetitions required to obtain an optimal solution with a specified success probability, and including classical overheads such as hyperparameter tuning. The complete quantum workflow needs to be benchmarked against state-of-the-art classical solvers under comparable solution-quality and success-probability requirements.

\subsection{Problem Instances for Simulation}
Given the lack of an existing library of Closest String or Substring problem instances at the simulation scale we were working with, we chose to generate our own.
We designed five methods to generate these problem instances based on different scenarios to incorporate the different sources of error that might occur in a real dataset.
The first version onvolved randomly generating a numpy array of integers in our alphabet (using \textit{k = 4}). The problem with this, is that the algorithm is amplifying the best solutions, but for randomly generated problem instances with non-trivial parameters, without a certain degree of luck no good solutions will exist. This is expected, as if there is no structural similarities to find, the best possible solutions will be the most repetitive pieces of noise. To what degree this is common will be purely random.
The next step was to generate reference string sets with a shared, planted substring and test if the algorithm would amplify and identify it. This was implemented by overwriting parts of the random problem instance with a randomly generated common substring of length \textit{n}. 
There were three other scenarios which modified the Planted Substring problem instance to attempt mimicking possible real-world scenarios created.
The first was allowing part of the substring to be clipped off the end of a reference string. This aimed to imitate the fact that, if you were looking for a piece of DNA in a gene segment, it is possible that the repeating piece was partially trimmed at the edge of the examined area and as such only partly included as part of the reference string. 
The second was the scenario where the substring encountered some mutation or "copy error" between reference strings in the reference string set. To do this, we set a fixed probability of each implanted substring character not being used as it was overwriting the random array.
The final problem instance type was the scenario where the substring had been both partially clipped and mutated. This was implemented by combining the above methods and was potentially the most realistic representation of a real set of reference strings that this might be applied to. 

For the Closest String simulations, we used the Planted and Clipped problem instances, with the substring length set as equal to the string length, such that a random array has randomised lengths of the same string overwrite each reference string.
For each problem size, we generated a set of 30 problem instances, then classically calculated and recorded the cost of each potential solution state for each problem instance and the output states (the counts outputted by measuring the state after $p$ iterations of the algorithm, after simulation for 1024 shots).
For $n=3,4,5,6 $ we simulated $p \in[0,30]$. 
By simulating the $p=0$ state for each problem instance, we find the value of $\sigma_{instance}$, where $\sigma_{instance}$ is the sampled standard deviation of the initial superposition state's costs for a particular instance, using this value to scale $\gamma$ to $\frac{\gamma}{\sigma}$. It was found that this scaling was unnecessary if not slightly detrimental after using the new deterministic values for $\gamma$, as such all data generated in the final run abstained from using it.
From this we could calculate the expectation values, standard deviations, variance for the costs, along with the measurement probability and mean expectation value of each subshell of the output state for each instance simulated.
We successfully did this for the Closest String with problem parameters set to $m=2$, $k=2$ and $n=3,4,5,6$, which are $15$, $18$, $23$, $26$ and qubit circuits respectively.
\section{Results and Discussion}

In Figure 5, showing state amplification after 30 iterations of the NV-QWOA compared to the original equal superposition, with states ranked left to right by optimality, it is clear that the optimal state is amplified significantly above any other. In Figure 5b, 5c and 5d, both the first and second states ranked are optimal states, with equal qualities. 
As such, we see the optimal state is amplified into becoming the state with the highest probability of measurement with probability greater than ~80$\%$ in every case shown, as desired. 
\begin{figure}[H]
    \centering
    \includegraphics[width=0.9\linewidth]{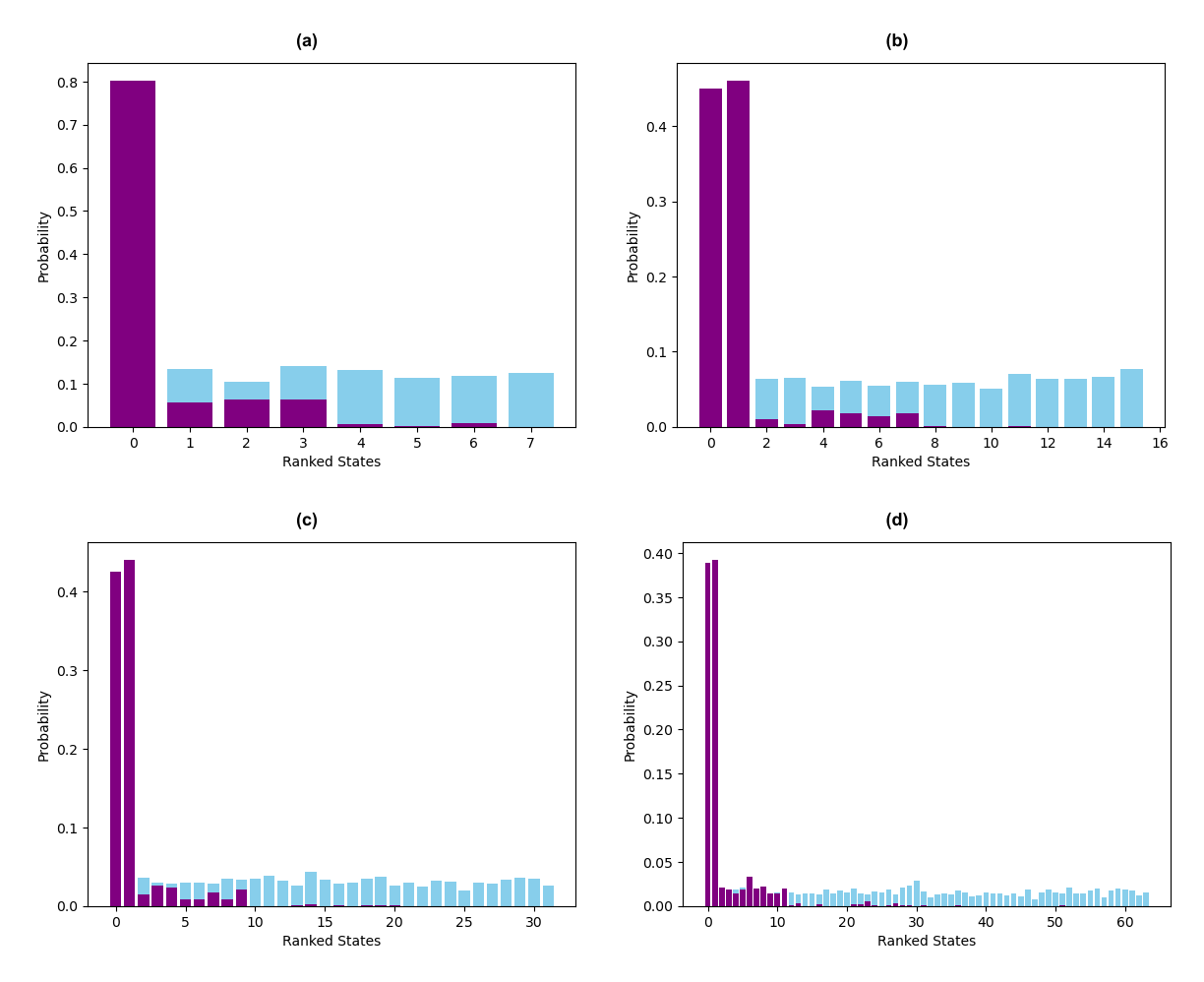}

    \caption{Shown are the $p=30$ amplified state (purple), overlayed with the initial p=0 superposition state (blue) for n=3,4,5,6 demarcated as a,b,c,d respectively.}
    \label{fig:placeholder}
\end{figure}
We see in Figure 6, that in all cases, subshell zero (the optimal solution set) is amplified to the largest probability as desired, at ~80$\%$  probability of observation, with subshell one retaining ~10-20$\%$ percent of the probability and all of the other subshells' probabilities converging to zero.
\begin{figure}[H]
    \centering
    \includegraphics[width=0.75\linewidth]{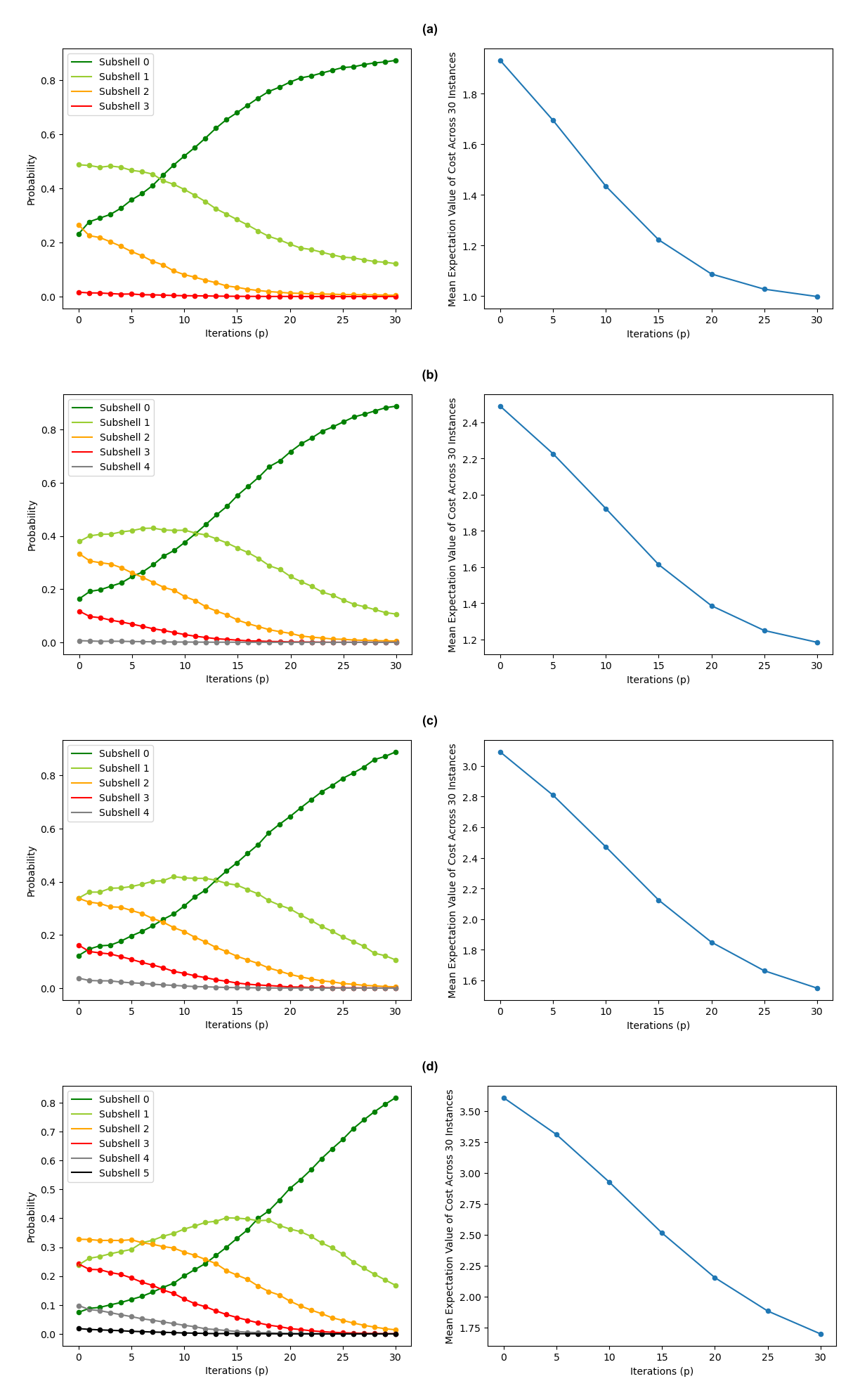}
    \caption{Column 1 shows the amplification of solutions in each subshell across $p$ iterations of the NV-QWOA, with subshell data averaged across 30 problem instances with input parameters: $k=2,m=2$, while Column 2 shows the mean expectation value of solution cost after $p$ iterations. Rows $a$ through $d$ represent $n=3,4,5,6$ respectively.}
\end{figure}
In Figure 7 there is a banding behaviour shown in the expectation value results, where the output expectation values of individual instances converge along the $n$ different paths to different final states at $p=30$. Most likely each band represents a different starting cost of subshell zero in that particular problem instance, as when the optimal solution is amplified beyond all others, this will overwhelmingly weight the expectation value to be close to the optimal solution cost. The most likely reason we don't see $n+1$ bands of instances is that the additional band would represent instances where the cost of subshell 0 is equal to $n$, meaning that every solution is contained in subshell zero. This would imply a complete lack of problem structure, such that every character of the best solution string would need to be changed, in every case for each reference string for them to match, and is actually not achievable with $m=2$, as such it doesn't appear here.

\begin{figure}[H]
    \centering
    \includegraphics[width=0.9\linewidth]{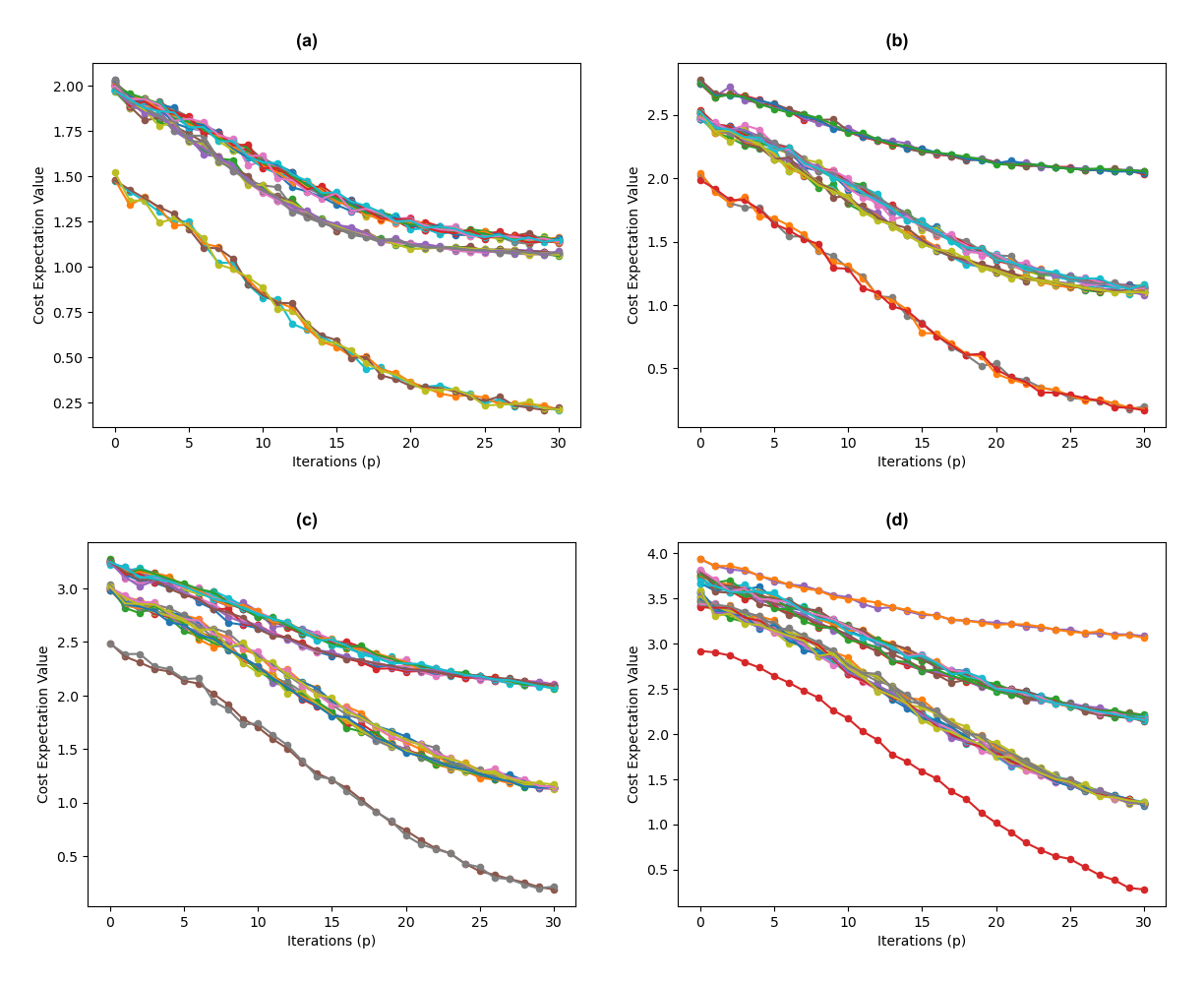}
    \caption{Shown are the the expectation values of cost for each of the 30 instances tested across iteration count, with $a$ through $d$ representing $n=3,4,5,6$ respectively parameters: $k=2,m=2$.}
    \label{fig:placeholder}
\end{figure}

\section{Conclusions}\label{Conclusions}

This work presents a quantum circuit framework for applying the Non-Variational Quantum Walk Optimization Algorithm (NV-QWOA) to the Closest String problem. The framework integrates reversible Hamming-distance evaluation, deterministic selection of the maximum distance across the reference strings, a cost-dependent phase shift, and quantum-walk mixing over a Hamming graph. Together, these components provide an explicit quantum gate-level implementation that encodes the optimization objective and redistributes measurement probability towards high-quality candidate strings.

Classical state-vector simulations of circuits using up to 28 qubits demonstrate the applicability of NV-QWOA to the Closest String problem. For the tested instances with string lengths $n=3,4,5,6$, two reference strings ($m=2$), and a binary alphabet ($k=2$), 30 iterations increased the total probability of measuring an optimal solution to approximately $80$–$90\%$. Starting from a uniform superposition over the $k^n$ candidate strings, the algorithm therefore concentrates a substantial fraction of the measurement probability on the optimal solution set. These results provide numerical evidence of effective solution amplification across the examined instances, although larger-scale studies are needed to establish how this performance scales with problem size.

The implementation requires $\mathcal{O}(nm\log k)$ qubits and has a circuit depth of $\mathcal{O}(pm\log^2 n+pn\log n)$, where $p$ is the number of NV-QWOA iterations. These resource estimates provide a basis for comparison with classical exact algorithms and for assessing future quantum hardware implementations. Establishing an end-to-end quantum speedup requires further analysis of the iteration count and measurement repetitions needed to achieve a specified success probability, together with any classical preprocessing and hyperparameter-tuning costs.
The structural relationship between the Closest String and Closest Substring problems, together with preliminary classical investigations, suggests that this framework could be extended to the Closest Substring problem. Such an extension would require incorporating substring selection into the candidate representation and objective evaluation. Although this possibility was not investigated in detail here, it offers a direction for future work.

\section{Acknowledgements}

We thank Tavis Bennett and Edric Matwiejew for valuable discussions on the initial project design and quantum-walk-based optimisation methodology. We also acknowledge the Pawsey Supercomputing Research Centre for substantial computational resources.

\bibliographystyle{apsrev4-2}
\bibliography{references}
\end{document}